\documentclass{osa-article}

\journal{oe}

\usepackage{bibentry}
\usepackage[utf8]{inputenc}
\usepackage{tikz}
\usepackage{svg}
\usepackage{comment}
\usepackage{gensymb}
\usetikzlibrary{matrix,shapes,arrows,positioning,chains}
\newcommand{\ignore}[1]{}
\newcommand{\nobibentry}[1]{{\let\nocite\ignore\bibentry{#1}}}

\newcommand{\da}[1]{\text{d}#1 }

\graphicspath{{figs_acoustic/}}

\articletype{Research Article}

\usepackage{lineno}
\linenumbers

\begin{document}

\title{Noise in Brillouin Based Information Storage}

\author{Oscar A. Nieves\authormark{1,*}, Matthew D. Arnold\authormark{1}, Miko\l{}aj K. Schmidt\authormark{2}, Michael J. Steel\authormark{2} and Christopher G. Poulton\authormark{1}}

\address{\authormark{1} School of Mathematical and Physical Sciences, University of Technology Sydney, 15 Broadway, Ultimo NSW 2007, Australia.
\authormark{2} Deparment of Physics and Astronomy, Macquarie University, North Ryde NSW 2109, Australia.}

\email{\authormark{*}oscar.a.nievesgonzalez@student.uts.edu.au}



\begin{abstract}
We theoretically and numerically study the efficiency of Brillouin-based opto-acoustic data storage in a photonic waveguide in the presence of thermal noise and laser phase noise. We compare the physics of the noise processes and how they affect different storage techniques, examining both amplitude and phase storage schemes. We investigate the effects of storage time and pulse properties on the quality of the retrieved signal, and find that phase storage is less sensitive to thermal noise than amplitude storage.
\end{abstract}

\section{Introduction}
\begin{figure}[h]
	\centering
	\includegraphics[width=1.0\textwidth]{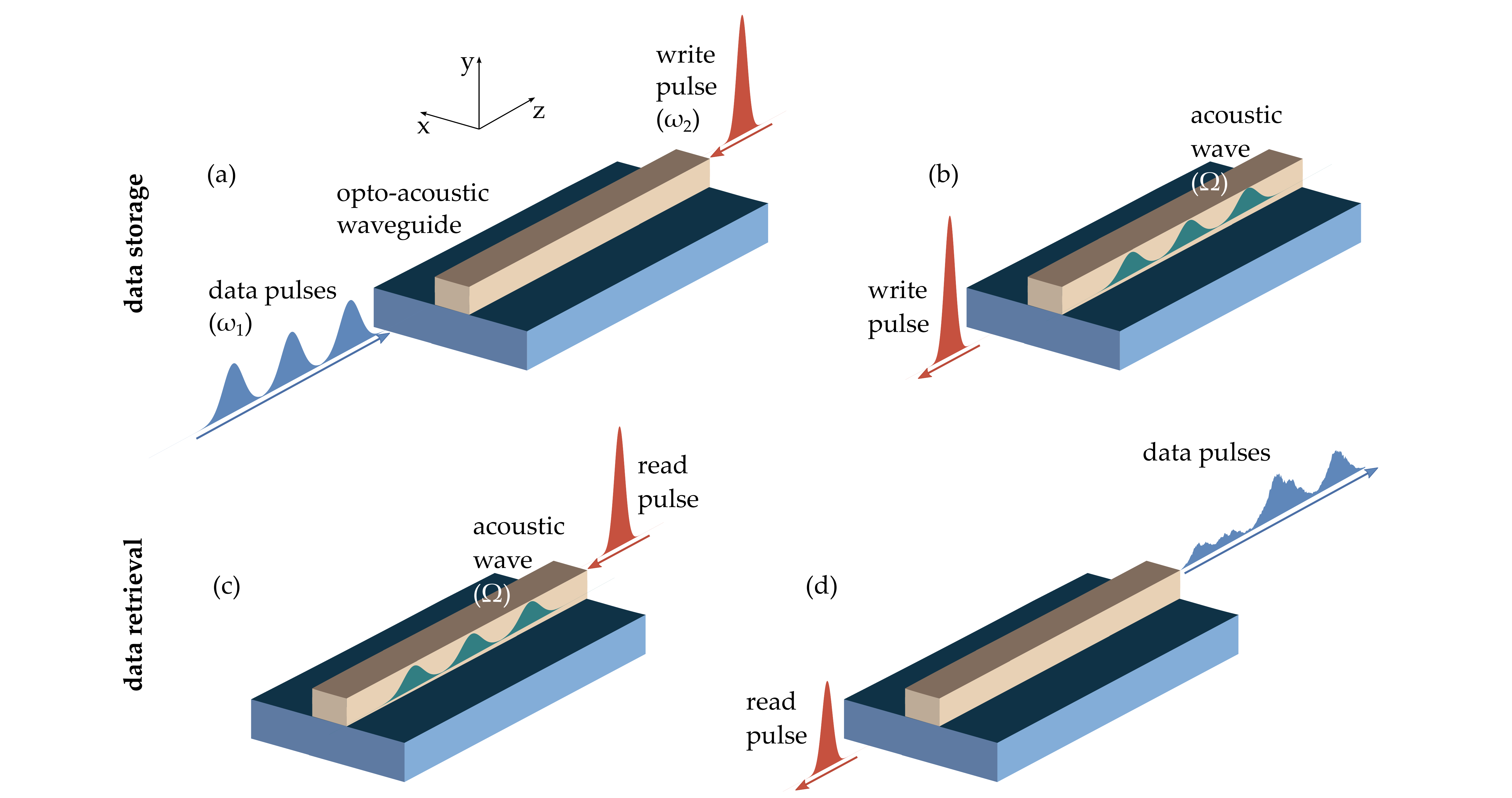}	
	\caption{Illustration of the opto-acoustic storage achieved via SBS, neglecting the effects of noise, and using amplitude shift-keying (amplitude storage). Storage process in (a) and (b): data pulses are depleted by the write-pulse, exciting an acoustic wave inside the waveguide, with some energy gained by the write-pulse. Retrieval process in (c) and (d): a read-pulse interacts with the acoustic wave, both become depleted and the energy is used to regenerate the original sequence of data pulses. The retrieval efficiency is limited by the acoustic lifetime of the phonons.}
	\label{fig:Storage_diagram}
\end{figure}

\noindent Stimulated Brillouin scattering (SBS) is an opto-acoustic process that results from the coherent interaction between two counter-propagating optical fields and an acoustic wave inside an optical waveguide~\cite{eggleton2019, kobyakov2010, pant2014, boyd2003, brillouin1922}. In backward SBS, a pump field at frequency $\omega_1$ is injected at one end of the waveguide, and a counter-propagating field at the Brillouin Stokes frequency $\omega_2 = \omega_1 - \Omega$, where $\Omega$ is the Brillouin frequency shift, is injected at the other;
the interaction of these two fields creates a strong acoustic wave at frequency
$\Omega$. When the Stokes power is comparable in size to the pump power, the pump energy can be completely depleted and some of it transferred to the acoustic field, along with its amplitude and phase information. A recent application of this effect is opto-acoustic memory storage
(see Fig.~\ref{fig:Storage_diagram}): an optical data pulse at the pump frequency interacts with an optical ``write'' pulse (at the Stokes frequency), creating an acoustic hologram where the original data is temporarily stored~\cite{zhu2007,kalosha2008,merklein2018}; an optical ''read'' pulse 
at the Stokes frequency can then be used to regenerate the original data pulse. Brillouin-based opto-acoustic information storage has been demonstrated experimentally in fibres~\cite{zhu2007} as well as more recently in on-chip 
experiments~\cite{kalosha2008,dong2015brillouin,merklein2017,merklein2018,merklein2020,stiller2020}, with storage times ranging up to 40 ns \cite{stiller2020}. Opto-acoustic storage, however, will inevitably be limited by noise, which 
exists because of the presence of thermal phonons in the waveguide, as well
as being an inherent feature of the lasers used in the data and read/write pulses.
At room temperatures, it is known that thermal noise significantly degrades the quality of Brillouin processes~\cite{boyd1990,gaeta1991,ferreira1994,kharel2016,behunin2018,nieves2021}, and it can be expected that it will also place limits on information \textit{retrieval efficiency}
in Brillouin storage experiments, measured as the amount of power that can be retrieved from the acoustic field after a certain time. This has been experimentally measured~\cite{merklein2020,zhu2007}, however while noise has been observed as a feature of these experiments, it is not yet clear how noise impacts the accuracy of the information retrieval. 
Furthermore, most studies (except, notably, \cite{merklein2017}) have focused on amplitude encoded storage, in which bits of information are stored in individual pulses. Alternatively, phase encoded storage may offer higher storage efficiency and be less sensitive to noise. A quantitative understanding of the impact of noise on the storage of both amplitude and phase information is needed for the further development of practical Brillouin-based storage devices.

In this paper, we apply our previous theoretical~\cite{nieves2021} and numerical~\cite{nieves2021num} models to the simulation of Brillouin storage 
in the presence of thermal noise. These studies focused on the noise properties of optical pulses in the case of SBS amplification, with the pump containing a lot more power than the input Stokes seed. Here, we investigate the effect of noise on both amplitude and phase storage of information, by using a small pump power and a large Stokes seed power. We quantify the retrieval efficiency and accuracy in the form of
the packet error rate (PER) of 8-bit sequences. We find that phase storage offers a significant improvement in the duration with which information can be stored without degradation due to thermal noise. We examine this effect in more detail by computing the effect of thermal noise on the amplitude and phase of a phase-encoded signal, and find that although the variance in amplitude and phase increases at the same rates, the phase information is more robust to noise in accordance with the additive-white-Gaussian-noise (AWGN) model of discrete communication theory. 

\section{Numerical simulation of noise}
\subsection{Brillouin coupled mode equations}
We consider a waveguide of length $L$ oriented along the $z$ coordinate
axis, in which an optical data stream propagates in the positive
$z$ direction. The interaction between the two optical fields and the acoustic field can be described by the coupled system
~\cite{nieves2021,nieves2021num}
\begin{align}
\label{eq:A1pde}\frac{\partial a_1}{\partial z} + \frac{1}{v} \frac{\partial a_1}{\partial t}  + \frac{1}{2}\alpha a_1 &= i\omega_1 Q_1 a_2 b^*,\\
\label{eq:A2pde}\frac{\partial a_2}{\partial z} - \frac{1}{v} \frac{\partial a_2}{\partial t}  - \frac{1}{2}\alpha a_2 &=  i\omega_2 Q_2 a_1 b,\\
\frac{\partial b}{\partial z} + \frac{1}{v_a} \frac{\partial b}{\partial t}  + \frac{1}{2}\alpha_{\text{ac}} b &= i\Omega Q_a a_1^* a_2 + \sqrt{k_B T \alpha_{\text{ac}}} R(z,t). \label{eq:bpde}
\end{align}
where $a_1(z,t)$ and $a_2(z,t)$ are the envelope fields
of the data pulse (pump) and read/write pulse (Stokes) respectively, which correspond to mode fields with frequency/wavenumber $(\omega_{1,2},k_{1,2})$.
$b(z,t)$ is the acoustic envelope field for the mode with 
frequency/wavenumber $(\Omega,q)$. 
All envelope fields have units of W$^{1/2}$.
In these equations the optical group velocity is given by $v>0$ and the acoustic group velocity is denoted $v_a>0$. $\alpha$ and $\alpha_{\text{ac}}$ are the optical and acoustic loss coefficients respectively (in units of m$^{-1}$). The coefficients $Q_{1,2,a}$ represent the coupling strength of the SBS interaction, which depend on the optical and acoustic modes of the waveguide~\cite{sturmberg2019}; from local conservation of energy, we have $Q_1 = -Q_2^*$ and $Q_a=Q_2^*$~\cite{wolff2015stimulated}. These parameters are related to the SBS gain parameter $g_0$ (with units of m$^{-1}$W$^{-1}$), via the expression $Q_2 = \sqrt{g_0\Gamma/(4v_a\omega_2\Omega)}$, where $\Gamma = v_a\alpha_{\text{ac}} = 1/\tau_a$ is the rate of decay of the phonons. In deriving Eq.~\eqref{eq:A1pde}$-$(\ref{eq:bpde}) we have assumed the phase matching conditions $\Omega = \omega_1 - \omega_2$ and $q = k_1-k_2$.

The thermal noise is represented by a zero-mean space-time white noise function, with auto-correlation function given by $\langle R(z,t)R(z',t')\rangle = \delta(z-z')\delta(t-t')$. The parameter $k_B T\alpha_{\text{ac}}$ is the strength of the thermal noise, which arises from the fluctuation-dissipation theorem~\cite{nieves2021}. The functions describing the data
stream and the read-write pulses are given by $a_{\text{data}}(t)$, $a_{\text{read}}(t)$ and  $a_{\text{write}}(t)$, which describe the values of the optical envelopes at $z=0$ (data) and $z=L$ (read/write). To incorporate phase noise from the lasers
used to generate these fields, the envelope fields at the ends of the waveguide are specified by stochastic boundary conditions, namely $a_1(0,t)=a_{\text{data}}(t) e^{i\phi_1(t)}$ for the data pulse and $a_2(L,t)=\left[a_{\text{read}}(t) + a_{\text{write}}(t)\right]e^{i\phi_2(t)}$. The random phase terms $\phi_{1,2}$ are statistically uncorrelated Brownian motions described by
\begin{equation}\label{eq:Random_phase}
    \phi_{j}(t) = \sqrt{2\pi\Delta\nu_L}\int_{-\infty}^{t}\da{W(t)},
\end{equation}
where $\da{W(t)}$ is a standard Wiener increment~\cite{horsthemke1984}, and $\Delta\nu_L$ is the laser's intrinsic linewidth~\cite{nieves2021num}. 

In non-storage SBS setups~\cite{nieves2021,nieves2021num} the pump has higher input power $P_{\text{p0}}$ than the Stokes $P_{\text{s0}}$ and thus behaves as an amplifier. However, in the case of Brillouin storage, we require $P_{\text{p0}} \ll P_{\text{s0}}$ so that the read/write pulse can completely deplete the pump, which contains the data to be stored. Optimum storage requires two conditions
on the read/write pulse. First, the read/write pulses must be at least as short in duration as the shortest data pulse.
Second, the pulse area for the read/write pulse, defined as~\cite{zhu2007,dong2015} 
\begin{equation}\label{eq:Area}
    \Theta_w = \sqrt{\frac{g_0 v}{8\tau_a}}\int_{-\infty}^{\infty} a_{\text{read/write}}(t)\da{t},
\end{equation}
must obey the condition $\Theta_w = (m+1/2)\pi$ where $m=0,1,2,...$. This is because once the data pulse is depleted completely, the transfer of energy reverses and the data pulse is regenerated~\cite{winful2013,dong2015,allen1987}. The dependence of the pulse-area on $\tau_a$ may seem deceptive at first, but since $g_0$ is also dependent on $\tau_a$ these two effects cancel out. 

The storage efficiency tells us how effective the storage system is, and may be defined as the ratio of the total output data power $|a_{\text{out}}(t)|^2$ to the total input data power $|a_{\text{data}}(t)|^2$~\cite{merklein2017}:
\begin{equation}\label{eq:S_efficiency}
    \eta_{\text{sto}} = \frac{\int_{0}^{\tau_{\text{data}}}\left\langle \left|a_{\text{out}}(t)\right|^2 \right\rangle \da{t} }{\int_{0}^{\tau_{\text{data}}}\left\langle \left|a_{\text{data}}(t)\right|^2 \right\rangle \da{t}},
\end{equation}
where $\langle X(t)\rangle = \frac{1}{N}\sum_{n=1}^{N} X_n(t)$ is the ensemble average of a function $X(t)$ over $N$ independent runs, and $\tau_{\text{data}}$ is the duration of the data train. Because the acoustic wave decays in time at a rate $1/\tau_a$, $\eta_{\text{sto}}$ decreases with longer storage times, reducing the efficiency in addition to the effects of noise. We define the storage time $\tau_{\text{sto}}$ as the temporal delay between the write and read pulses.

\subsection{Solution to the coupled mode equations}
We numerically solve Eq.~\eqref{eq:A1pde}$-$(\ref{eq:bpde}) following the numerical method described in~\cite{nieves2021num}. In brief, we note that the propagation distance of the acoustic wave over the time-scale of the SBS interaction is very small~\cite{boyd1990}. We therefore apply the limit $\partial_z b \rightarrow 0$ in Eq.~\eqref{eq:bpde}, and solve the equation in time via the Green function method~\cite{nieves2021}. The thermal noise function is simulated starting at thermal equilibrium at $t=0$, by drawing random numbers from a steady-state probability distribution corresponding to an Ornstein-Uhlenbeck process~\cite{uhlenbeck1930,nieves2021num}. We then substitute the formal solution $b(z,t)$ into Eq.~\eqref{eq:A1pde} and (\ref{eq:A2pde}), and integrate the two equations using a two-step iteractive method: first, the optical fields are translated in space by a distance $\Delta z = v\Delta t$ while the acoustic field remains stationary~\cite{nieves2021num}. Then, the optical fields at each point in space $z$ are updated by integrating the equations in time with an Euler-Mayurama scheme~\cite{kloeden1992}. This two-step process is iterated over multiple time steps to simulate the dynamics of the input pulses.

\begin{table}[h]
\centering
\begin{tabular}{||c c || c c||} 
 \hline
Parameter & Value & Parameter & Value \\ [0.5ex] 
 \hline\hline
Waveguide length $L$ & 30 cm & Peak read/write power & 0.5$-$10.5 W\\
Waveguide temperature $T$ & 300 K & Data packets & 256\\
Refractive index $n$ & 2.44 & Data stream duration $\tau_{\text{data}}$ & 2.44 ns\\
Acoustic velocity $v_a$ & 2500 m/s & Bit duration $\tau_{\text{bit}}$ & 300 ps \\
Acoustic lifetime $\tau_a$ & 10.2 ns & Data pulse width $\tau_1$ & 150 ps \\
Brillouin shift $\Omega/2\pi$ & 7.8 GHz & Read pulse width $\tau_2$ & 100 ps\\
Brillouin gain parameter $g_0$ & 411 m$^{-1}$W$^{-1}$ & Grid size (space) $N_z$ & 800\\
Optical wavelength $\lambda$ & 1550 nm  & Grid size (time) $N_t$ & 2797\\ 
Laser linewidth $\Delta\nu_L$ & 100 kHz & Step-size $\Delta z$ & 375 $\mu$m\\ 
Peak data power & 10 mW & Step-size $\Delta t$ & 3.05 ps\\ 
Optical loss $\alpha$ & 0.1 dB/cm & & \\[1ex]
 \hline
\end{tabular}
\caption{Parameters used in this study. Physical parameters correspond to a chalcogenide waveguide \cite{xie2019}.}
\label{table:tab1}
\end{table}

\noindent For our simulations we use the parameters summarized in Table~\ref{table:tab1}. We assume a high gain chalcogenide waveguide, of the type used in previous SBS experiments~\cite{xie2019}. We store individual 8-bit packets one at a time, consisting of all 256 possible unique 8-bit sequences. Each packet corresponds to 8 individual pulses in the amplitude storage case, and 4 phases in the phase storage case, as shown in Fig.~\ref{fig:Encoding_diagram}.  

It should be noted that the model used here (Eq.~\eqref{eq:A1pde}$-$(\ref{eq:bpde})) makes the assumption of the slowly-varying envelope approximation (SVEA) and rotating-wave approximation (RWA)~\cite{nieves2021,wolff2015stimulated}. Because the optical frequencies used in SBS experiments are typically in the range of a few hundred THz~\cite{eggleton2019,wolff2021}, this means that the pulses simulated must be wider than a few hundred picoseconds so that these approximations remain valid~\cite{piotrowski2021}. Therefore, we limit our simulations to pulses no shorter than 100 ps, for both for the data and read/write cases. We have chosen to focus on the storage of 8-bit sequences because they can fit inside a 30 cm chalcogenide waveguide without being too short to violate the SVEA and RWA, since the acoustic pulses generated from 300 ps optical bits are approximately 3.7 cm in length. 

\subsection{Data encoding}
\begin{figure}[h]
	\centering
	\includegraphics[width=1.0\textwidth]{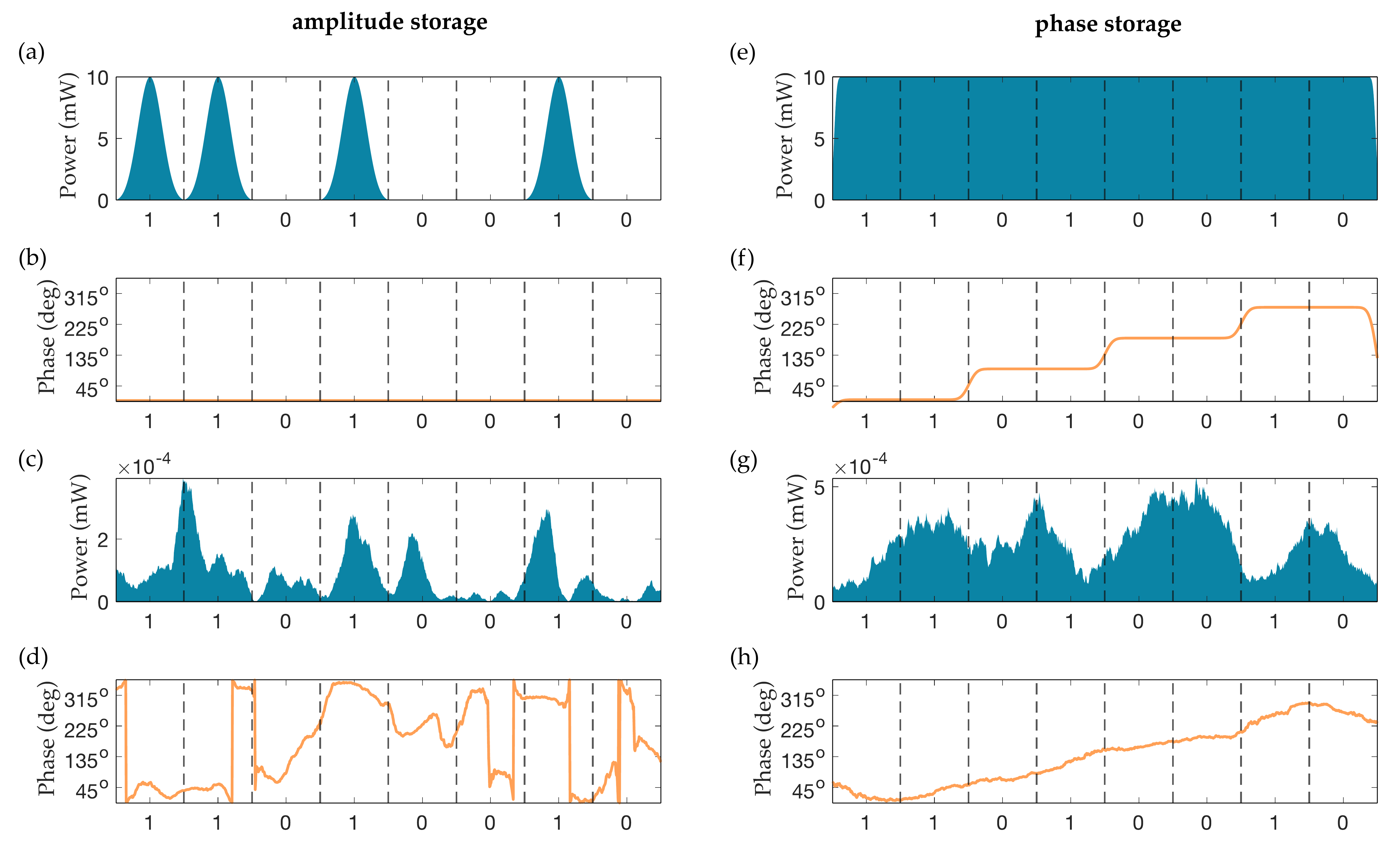}
	\caption{Illustration of how the 8-bit sequence $\mathbf{x} = 11010010$ is encoded in the numerical simulation. Amplitude storage: (a)$-$(b) input power and phase, (c)$-$(d) output power and phase after some storage time. Phase storage: (e)$-$(f) input power and phase, (g)$-$(h) output power and phase after some storage time. Here we use thermal noise at $T=300$ K.}
	\label{fig:Encoding_diagram}
\end{figure}
\begin{figure}[h]
	\centering
	\includegraphics[width=\textwidth]{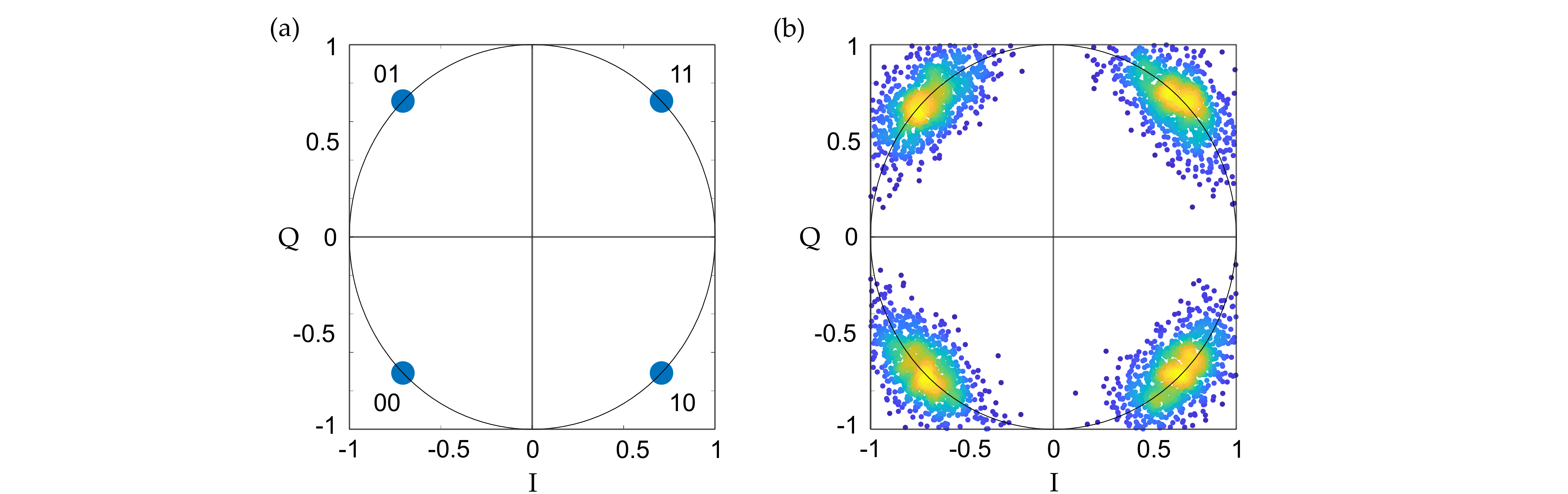}
	\caption{Constellation diagrams used in phase storage. These plots are done on the complex plane, where the modulus of each point corresponds to the normalized power of the signal, and the argument corresponds to the phase of the signal. (a) ideal system: each point lies at exactly 45$\degree$ from the horizontal line in either direction, and on the circle of radius 1. (b) Effect of noise in the detected signal: amplitude noise shifts the points in the radial direction, while phase noise shifts the points in the angular direction, and the color scheme shows the regions of higher or lower density of points.}
	\label{fig:QPSK_diagrams}
\end{figure}
\noindent The two storage schemes used in this paper --- amplitude and phase storage --- are summarized in Fig.~\ref{fig:Encoding_diagram}. Each bit is defined as having a duration $\tau_{\text{bit}}$. For amplitude storage, we encode ones into Gaussian pulses of the same peak power $P_{\text{p0}}$, while zeros are represented by gaps of duration $\tau_{\text{bit}}$. For phase storage, we use the same scheme as in quadrature phase-shift keying (QPSK) with gray coding~\cite{faruque2017}, where bit pairs are assigned a unique phase, namely $11 = 45\degree$, $01 = 135\degree$, $00 = 225\degree$ and $10 = 315\degree$.  For a given input information packet, we quantify the retrieval accuracy in both storage schemes via the packet error rate (PER). This is similar to the bit error rate (BER) of a binary stream, except that we count correct 8-bit packets as opposed to counting individual correct bits. Therefore, the PER is the ratio of correctly retrieved packets with respect to the input data, and has a value $0 \leq\text{PER}\leq 1$. 

In the amplitude storage case, we encode bits=1 into Gaussian pulses of full-width at half-maximum (FWHM) $\tau_1$, while bits=0 are represented by gaps of duration $\tau_{\text{bit}}$ in the data sequence. By default, we choose all pulses to have a phase of 0$\degree$. In the retrieval stage, the output data power $|a_{\text{out}}(t)|^2$ is separated into equal intervals of length $\tau_{\text{bit}}$. We use a dynamic threshold technique: initially, a threshold power $P_{\text{thresh}}$ is set. Then, we record the output power at the center of the bit period, and a single bit is read as 0 if $|a_{\text{out}}^{(n)}(t)|^2<P_{\text{thresh}}$ and as 1 if $|a_{\text{out}}^{(n)}(t)|^2\geq P_{\text{thresh}}$~\cite{walsh2005}, for all 256 data packets. This process is repeated for different values of $P_{\text{thresh}}$ until the total PER has been minimized.

In the phase storage case, we encode bit-pairs into 4 different phases, as shown in Fig.~\ref{fig:QPSK_diagrams}(a), as part of an analytic smooth rectangular pulse (ASR)~\cite{granot2019} with FWHM $\tau_{\text{data}}$. We then apply a Gaussian filter to remove non-physical discontinuities;
to emulate the effect of the modulators used in SBS experiments
we have chosen a filter width of 5 GHz, resulting 
in a smooth phase transition between the bit-pairs (see Fig.~\ref{fig:Encoding_diagram}(f)). In the retrieval stage, $|a_{\text{out}}(t)|^2$ is separated into equal intervals of length $2\tau_{\text{bit}}$. We extract the phase from the amplified output signal as $\phi_{\text{out}}(t) = \tan^{-1}\left(\text{Im}\left[a_{\text{out}}(t)\right]/\text{Re}\left[a_{\text{out}}(t)\right]\right)$. As with the case of amplitude storage, we assume ideal detector conditions and record the phase at the center of the $2\tau_{\text{bit}}$ period, such that if $0\degree<\phi_{\text{out}}(t) <90\degree$, we read the output bit-pair as 11, if $90\degree<\phi_{\text{out}}(t) <180\degree$ we read the bit-pair as 01 and so on. In phase storage, we use another measure of signal integrity based on the constellation diagram data: let $z_{\text{out}}^{(m)}$ represent a single point on the constellation diagram corresponding to the $m$th bit-pair in the binary sequence (such as 00), with magnitude $\left|z_{\text{out}}^{(m)}\right|$ and phase $\phi_{z}^{\text{m}}$. The variances in each can be found via
\begin{equation}
    \text{Var}\left[\left|z_{\text{out}}^{(m)}\right|\right] = \left\langle \left|z_{\text{out}}^{(m)}\right|^2 \right\rangle - \left\langle \left|z_{\text{out}}^{(m)}\right| \right\rangle^2,\quad
    \text{Var}\left[\phi_{\text{out}}^{(m)}\right]  = \left\langle \left(\phi_{\text{out}}^{(m)}\right)^2 \right\rangle - \left\langle \phi_{\text{out}}^{(m)} \right\rangle^2.
\end{equation}

\section{Results and Discussion}
\subsection{Effect of thermal noise}
\begin{figure}[h]
	\centering
	\includegraphics[width=1.0\textwidth]{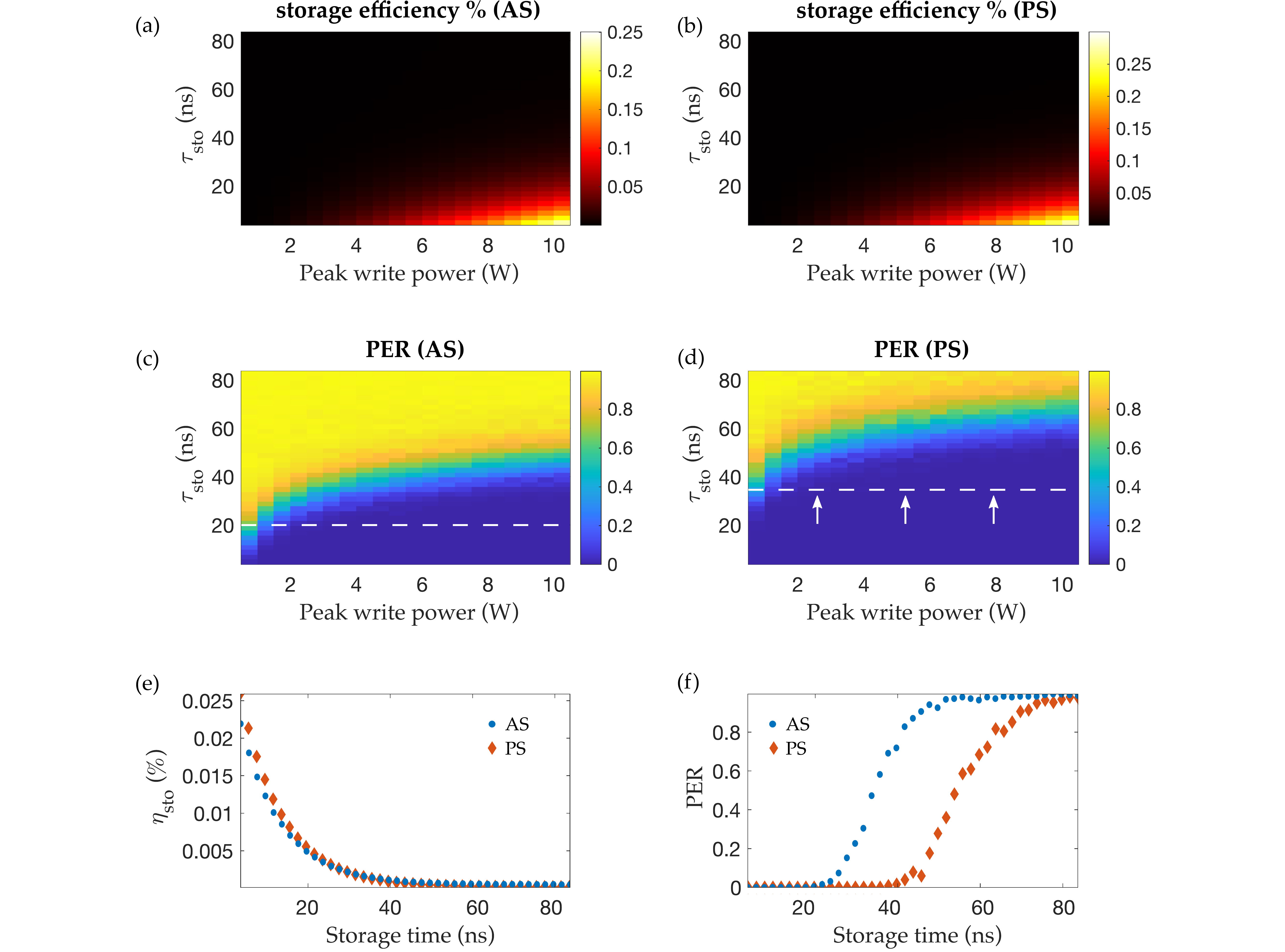}
	\caption{Simulations of Brillouin storage with thermal noise only, for varying read/write pulse peak power and different storage times in amplitude storage and phase storage. Panels (a) and (b) show the storage efficiency of the two schemes, (c) and (d) the packet error rates, (e) and (f) illustrate the storage efficiency and PER at 3 W peak write power. The dashed horizontal line in (c) and (d) indicates the $\tau_{\text{sto}}$ at which data packet errors begin to occur, and the arrows in (d) show that this threshold is pushed further back in time.}
	\label{fig:Long_BER1}
\end{figure}

\begin{figure}[h!]
    \centering
    \includegraphics[width=\textwidth]{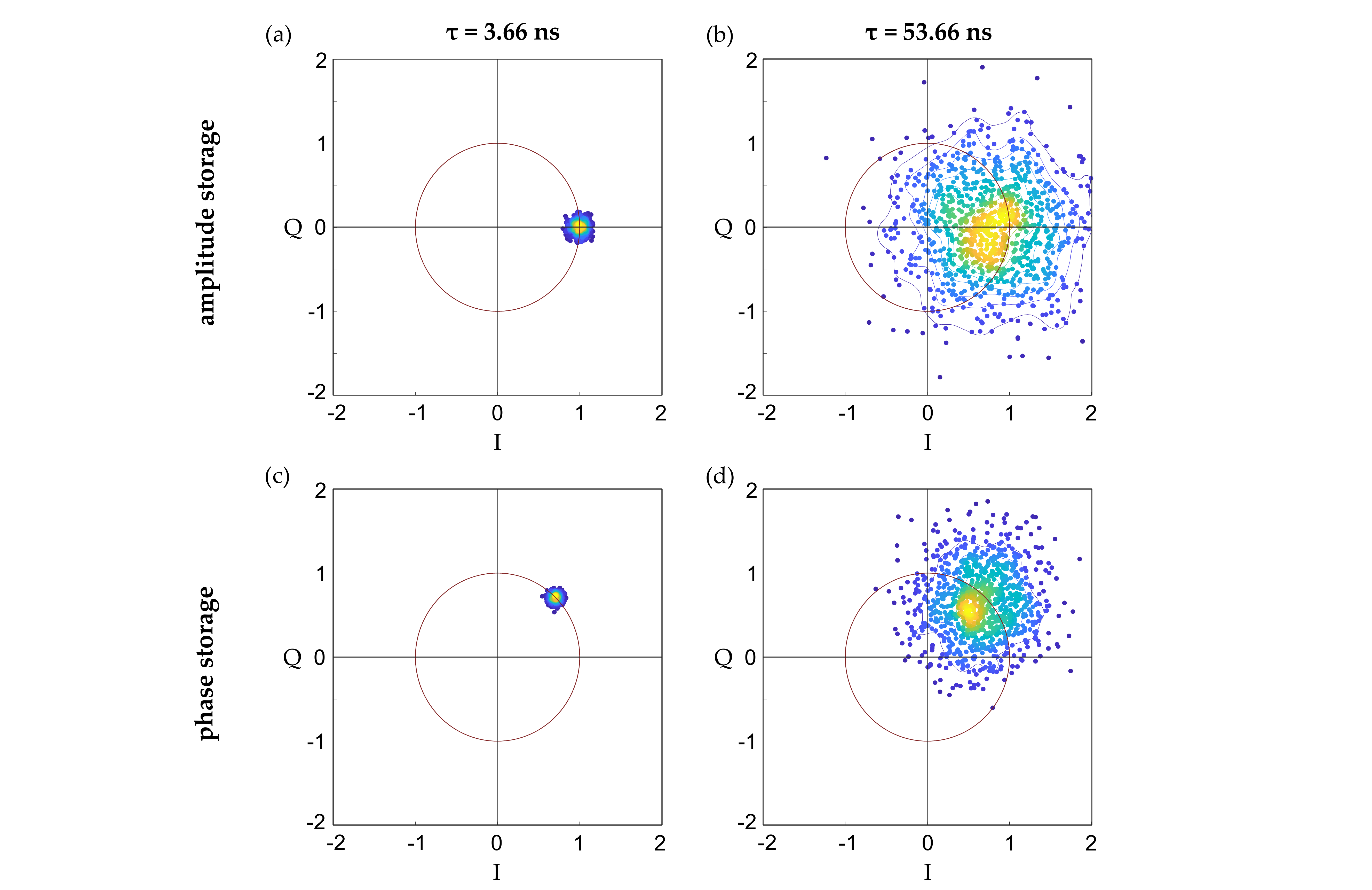}
    \caption{Constellation diagrams for the thermal noise only case, at 3 W peak write power. (a) and (b) show the amplitude storage plots at two storage times (3.66 ns is the minimum storage time achievable in this configuration) for a binary bit 1, while (c) and (d) show the phase storage plots for a binary bit pair 11.}
    \label{fig:Const1}
\end{figure}

\begin{figure}[h!]
    \centering
    \includegraphics[width=\textwidth]{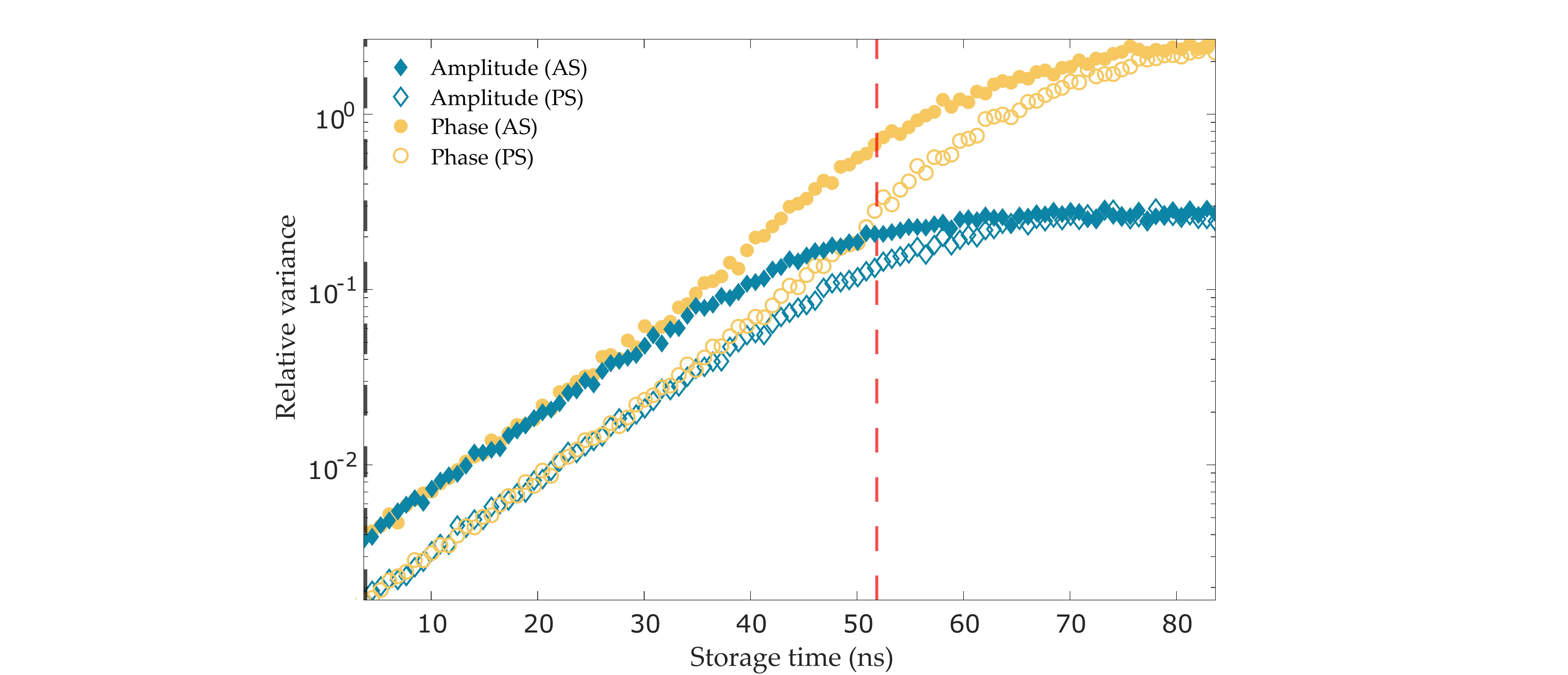}
    \caption{Relative variance in the amplitude and phase of the constellation data in Fig.~\ref{fig:Const1} as a function of storage time. The dashed lines mark the storage times corresponding to the constellation diagrams in Fig.~\ref{fig:Const1}.}
    \label{fig:Var1}
\end{figure}

\noindent We investigate the effect of varying the read/write pulse peak power between 0.5$-$10.5 W, while maintaining a read/write pulse width of 100 ps. We include thermal noise into the waveguide at 300 K, but neglect the input laser phase noise by setting the laser linewidth $\Delta\nu_L$ to zero. In the amplitude storage case, we use Gaussian data pulses with 10 mW peak power and pulse width 150 ps, while in the phase storage case we use a rectangular pulse of duration 2.44 ns, and phase intervals of duration 600 ps. Each bit of optical information is 300 ps in length for both amplitude and phase storage. The results of the simulations are shown in Fig.~\ref{fig:Long_BER1}. First, we observe in Fig.~\ref{fig:Long_BER1}(a) and (b) that the storage efficiency in both storage schemes is higher as we increase the peak write pulse power, as the pulse area is lower than the optimum value given by Eq.~\eqref{eq:Area}. Second, we see in Fig.~\ref{fig:Long_BER1}(c) and (d) that lower peak write powers lead to increased PER, thus reducing the maximum storage time achievable in both encoding schemes. This occurs because at lower peak read/write powers --- which also corresponds to the lower storage efficiency regime --- the coherent output data field is less distinguishable from the amplified spontaneous noise arising from the interaction between the read/write pulse and the thermal background fluctuations in the waveguide~\cite{nieves2021,nieves2021num}. Consequently, this increases the probability of retrieving the wrong data sequence at the waveguide output ($z=L$).

Similarly, we see an increase in PER in both AS and PS schemes with increasing $\tau_{\text{sto}}$, as shown in Fig.~\ref{fig:Long_BER1}(c) and (d). This occurs because in the time between the write and read-process, the acoustic wave containing the stored information decays at a rate $1/\tau_a$. As $\tau_{\text{sto}}$ increases, the acoustic wave gets closer to the background thermal noise, increasing the fluctuations in the retrieved data field  during the read-process. This effect is more clearly illustrated in Fig.~\ref{fig:Const1}, where the constellation points spread out in both radial and angular directions, indicating an increase in both amplitude and phase noise in the retrieved data field. In Fig.~\ref{fig:Var1} we see that the rate of increase in the variance of the phase and amplitude of each encoding scheme is the same for the first 40 nanoseconds. However, in Fig.~\ref{fig:Long_BER1}(f) we see that the PER in the phase storage case begins to increase at longer $\tau_{\text{sto}}$ compared to the amplitude storage case, suggesting that phase-encoded data is more robust to thermal noise and hence allows for longer storage times. This occurs because the phase variations are primarily constrained to a single quadrant on the constellation plots (as shown in Fig.~\ref{fig:Const1}), whereas the amplitude variations reach the noise floor more rapidly. Consequently, the probability of detecting the wrong bit of information in the amplitude is higher compared to the phase. This is the same observation that would be expected from the additive white Gaussian noise (AWGN) model in discrete communication theory, where the probability of bit error for phase-shift-keying (PSK) is lower than for amplitude-shift-keying (ASK)~\cite{smith2012,bala2021}. In addition, phase encoding allows the transfer of more bits per symbol, and this means that the pulse-size constraints imposed by the SVEA and RWA can be further relaxed.

\subsection{Effect of laser phase noise}
\begin{figure}[h!]
	\centering
	\includegraphics[width=1.0\textwidth]{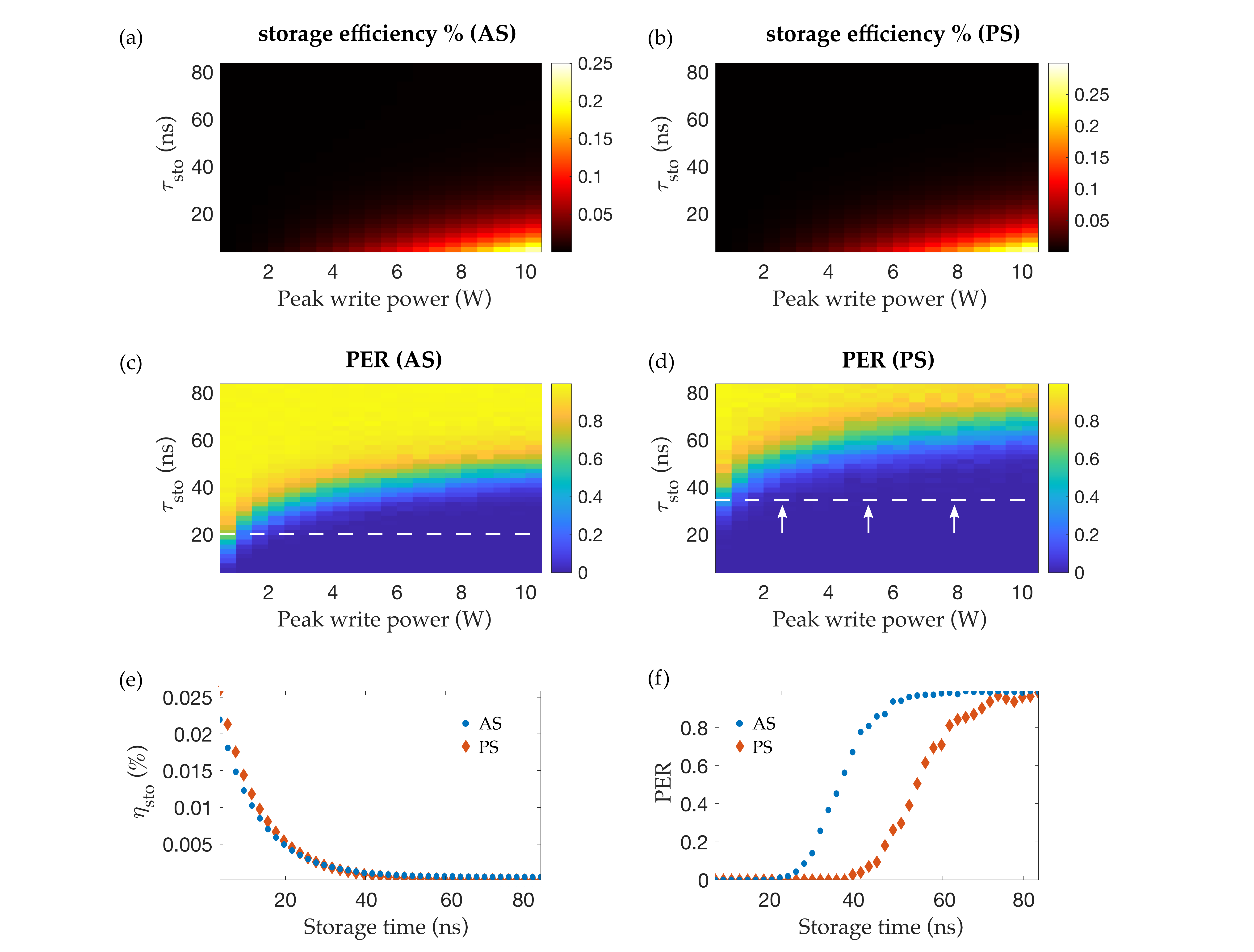}
	\caption{Simulations of Brillouin storage with both thermal and laser phase noise, for varying read/write pulse peak power and different storage times in amplitude storage and phase storage. Panels (a) and (b) show the storage efficiency of the two schemes, (c) and (d) the packet error rates, (e) and (f) illustrate the storage efficiency and PER at 3 W peak write power. The dashed horizontal line in (c) and (d) indicates the $\tau_{\text{sto}}$ at which data packet errors begin to occur, and the arrows in (d) show that this threshold is pushed further back in time.}
	\label{fig:Long_BER2}
\end{figure}

\begin{figure}[h!]
    \centering
    \includegraphics[width=\textwidth]{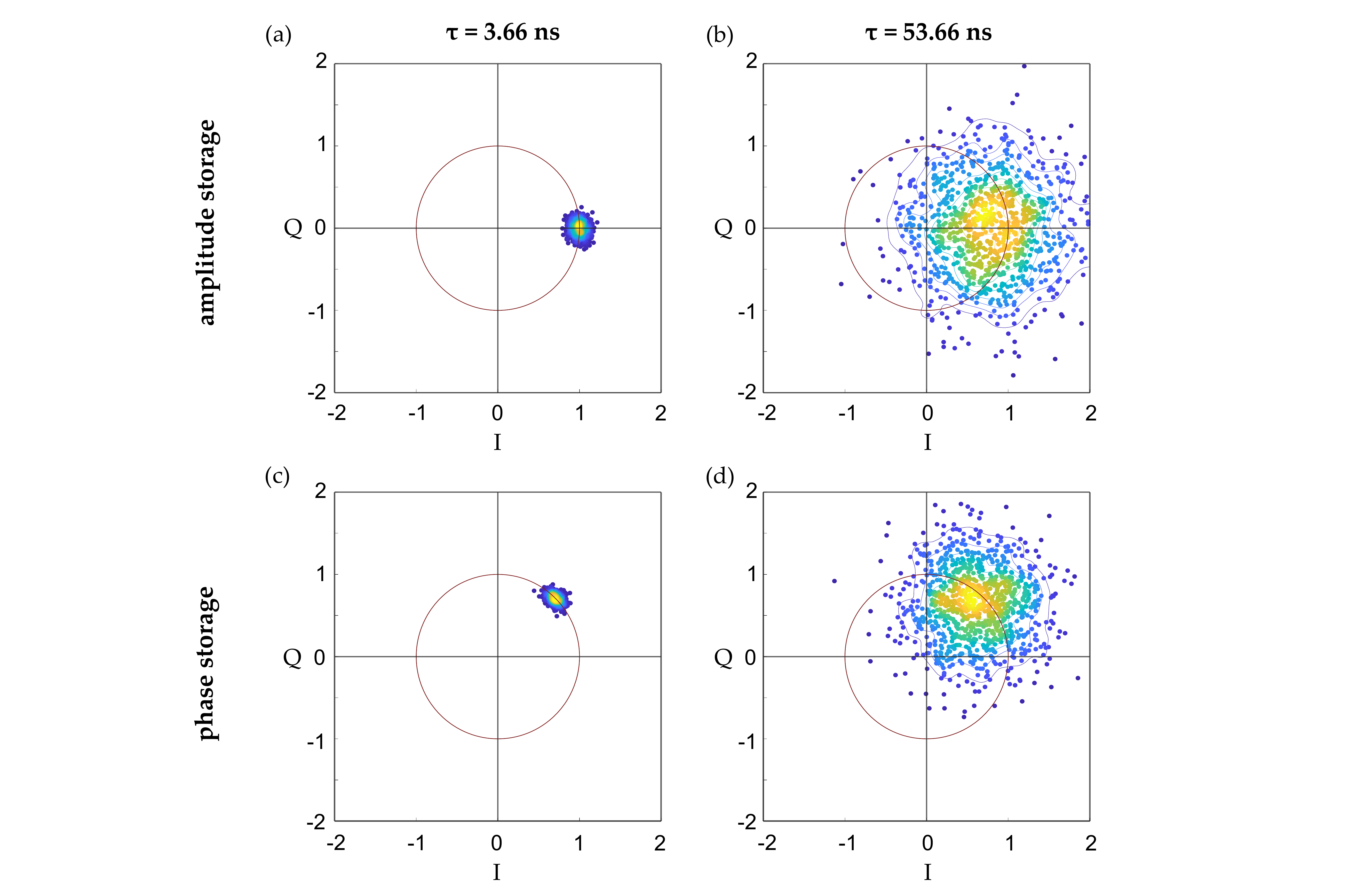}
    \caption{Constellation diagrams for the thermal noise and laser phase noise case, at 3 W peak write power. (a) and (b) show the amplitude storage plots at two storage times (3.66 ns is the minimum storage time achievable in this configuration) for a binary bit 1, while (c) and (d) show the phase storage plots for a binary bit pair 11.}
    \label{fig:Const2}
\end{figure}

\begin{figure}[h!]
    \centering
    \includegraphics[width=\textwidth]{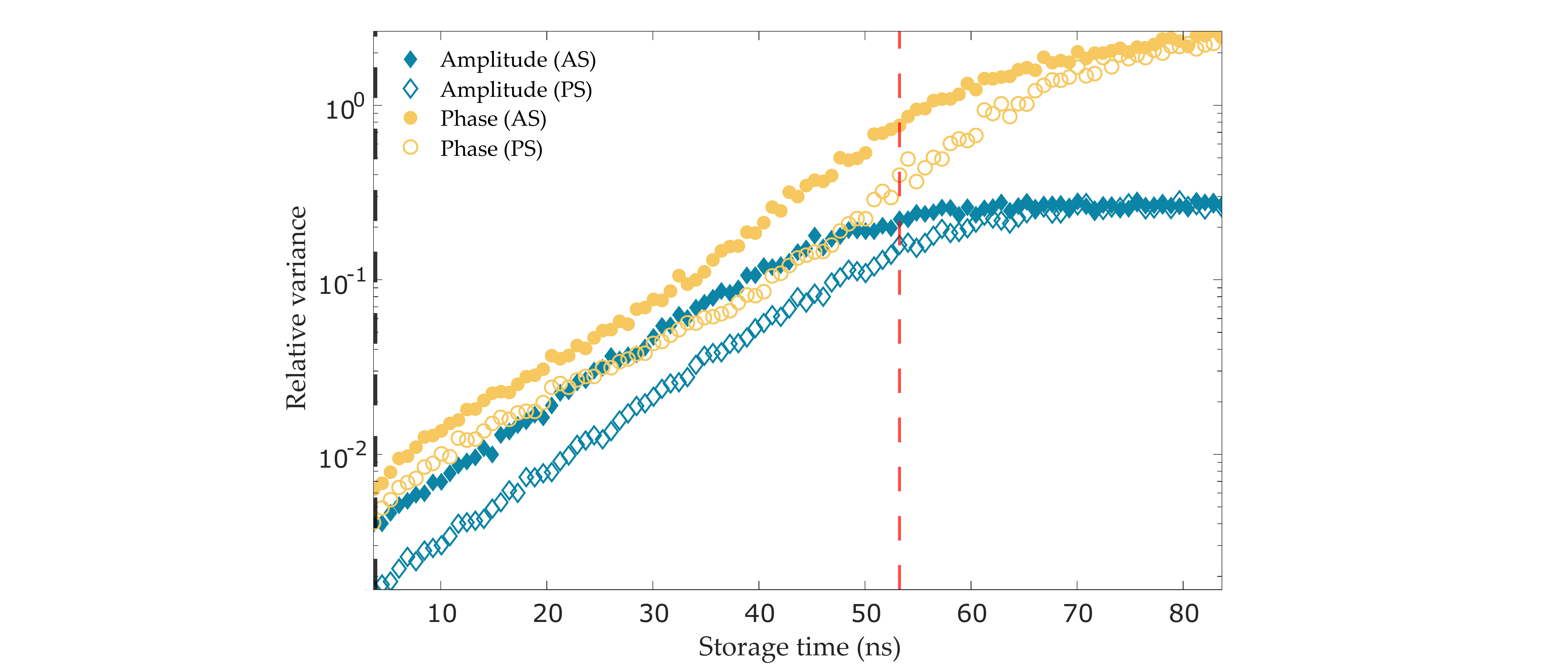}
    \caption{Relative variance in the amplitude and phase of the constellation data in Fig.~\ref{fig:Const2} as a function of storage time. The dashed lines mark the storage times corresponding to the constellation diagrams in Fig.~\ref{fig:Const2}.}
    \label{fig:Var2}
\end{figure}

\noindent We now investigate the effect of adding input laser phase noise, at a linewidth of 100 kHz, with the same pulse and thermal noise parameters as before. We generate the phase noise in the data pulses and read/write pulses independently so they are statistically uncorrelated, but have the same mean and variance properties. Fig.~\ref{fig:Long_BER2} shows the results for storage efficiency and PER for both encoding schemes. The plots in Fig.~\ref{fig:Long_BER2}(a)$-$(d) look very similar to Fig.~\ref{fig:Long_BER1}(a)$-$(d), indicating that laser phase noise does not have a significant impact on the storage efficiency or PER. This is consistent with previous work on SBS noise in short pulses~\cite{nieves2021num}, where it was found that laser phase noise only has a significant impact on the SBS process when the laser coherence time ($\tau_{\text{coh}} = 1/\pi\Delta\nu_L$) is comparable in magnitude to the SBS interaction time. Next, the constellation diagram results in Fig.~\ref{fig:Const2} reveal that the input laser phase noise broadens the variance in the phase of the retrieved data field, making the distribution of constellation points slightly more elliptical compared to the thermal noise only case. This is also shown in Fig.~\ref{fig:Var2}, where the rates of increase for the amplitude and phase variances still remain approximately the same for the first 40 nanoseconds of storage time. This occurs because at this regime of powers for the data pulses, the SBS process acts as a linear amplifier, hence the total noise in the retrieved data is a linear combination of the waveguide thermal noise and the laser phase noise from the inputs. 

\section{Conclusion}
We have numerically simulated the Brillouin storage of different data packets with thermal and laser noise, using amplitude storage and phase storage techniques in a photonic waveguide. Through these computer simulations, we have shown that phase encoded storage allows for longer storage times than amplitude encoded storage. This is because phase encoding is more robust to noise than amplitude encoding, in accordance with the additive-white-Gaussian-noise model of discrete communications theory~\cite{smith2012}. It is therefore possible to increase Brillouin storage time by encoding information into the phase of the data field, without having to change the waveguide or laser properties. Furthermore, because phase storage techniques can encode more bits per symbol than amplitude storage~\cite{faruque2017,smith2012}, the pulse size constraints imposed by the SVEA and RWA~\cite{piotrowski2021} in these mathematical models can be further relaxed when using phase encoding techniques.

\begin{backmatter}
\section*{Disclosures}
The authors declare no conflicts of interest.

\section*{Acknowledgments}
The authors acknowledge funding from the Australian Research Council (ARC) (Discovery Projects DP160101691, DP200101893), the Macquarie University Research Fellowship Scheme (MQRF0001036) and the UTS Australian Government Research Training Program Scholarship (00099F).  Part of the numerical calculations were performed on the UTS Interactive High Performance Computing (iHPC) facility. 

\section*{Data availability Statement} 
\noindent Data underlying the results presented in this paper are not publicly available at this time but may be obtained from the authors upon reasonable request.
\end{backmatter}

\bibliography{acoustic_bib}

\end{document}